\documentclass[aps,preprint]{revtex4}

\def\BibTeX{{\rm B\kern-.05em{\sc i\kern-.025em b}\kern-.T 08em\kern-.1667em\lower.7ex\hbox{E}\kern-.125emX}}
\usepackage{graphicx}
\usepackage{amsmath}  
\usepackage[utf8]{inputenc}
\usepackage[english]{babel}
\usepackage{amsfonts}
\usepackage{amssymb}

\begin{document}

\title{Negative permeability in magnetostatics and its experimental demonstration}

\author{Rosa Mach-Batlle$^1$, Albert Parra$^1$, Jordi Prat-Camps$^2$, Sergi Laut$^1$, Carles Navau$^1$, and Alvaro Sanchez$^{1,*}$}

\affiliation{$^1$ Departament de F\'{\i}sica, Universitat Aut\`onoma de Barcelona, 08193 Bellaterra,
Barcelona, Catalonia, Spain\\$^2$ Institute for Quantum Optics and Quantum Information of the Austrian Academy of Sciences, A-6020 Innsbruck, Austria\\
Institute for Theoretical Physics, University of Innsbruck, A-6020 Innsbruck, Austria\\
$^{*}$ alvar.sanchez@uab.cat}

\begin{abstract}
The control of magnetic fields, essential for our science and technology, is currently achieved by magnetic materials with positive permeability, including ferromagnetic, paramagnetic, and diamagnetic types. Here we
introduce materials with negative static permeability as a new paradigm for manipulating magnetic fields. As a first step, we extend the solutions of Maxwell magnetostatic equations to include negative-permeability values. The understanding of these new solutions allow us to devise a negative-permeability material as a suitably tailored set of currents arranged in space, overcoming the fact that passive materials with negative permeability do no exist in magnetostatics. 
We confirm the theory by experimentally creating a spherical shell that emulates a negative-permeability material in a uniform magnetic field. Our results open new possibilities for creating and manipulating magnetic fields, which can be useful for practical applications.

\end{abstract}

\maketitle

\section{Introduction}
Controlling magnetic fields is fundamental in both science and technology. Magnetic memories for computers, turbines for energy generation, motors for delivering power, medical techniques based on magnetic fields for treatment and diagnosis, they are all based on detailed spatial distributions of magnetic fields. Magnetic materials are the conventional tools to shape magnetic fields. This shaping of steady fields is governed by the laws of magnetostatics. One of the most important consequences of such laws is that fields decay from the sources, typically as dipoles, in contrast to the long-distance propagation of time-dependent electromagnetic waves.

The progress in the science of magnetic materials has been enormous in the last decades. Non-linear magnetic materials have been developed to attain complex behaviors including hysteresis and history effects, which are exploited in many actual technologies, like magnetic memories based on remanent magnetization states. 
In this work we focus on linear magnetic materials, those that have a magnetization directly proportional to the field. They have also experienced very important recent developments. One particularly active line of research is the application of linear magnetic materials as building blocks to construct magnetic metamaterials. 
Following the discovery of transformation optics technique and the development of metamaterials for the control of electromagnetic waves 
\cite{controlling,TO_review,leonhardtbook}, 
magnetic metamaterials have recently been introduced.
They have led to interesting new properties and devices for controlling magnetostatic fields \cite{wood,jung}, including magnetic cloaks 
\cite{wood,antimagnet,narayana,science,accloak,carpet_static,genenko,genenko2,hecloak}, magnetic concentrators \cite{concentrator,bjork,50T,dc_magnetic_conc,expconcentrator,liu}, and other novel magnetic phenomena \cite{hose,wormhole,transform_sunhe,jung}.   
The use of metamaterials is particularly attractive in magnetostatics, because of at least two properties. First, in the static case electric and magnetic fields decouple, so controlling magnetic fields requires only dealing with permeabilities \cite{wood}. Second, natural materials exist with extreme permeability values, such as $\mu\to 0$ and $\mu \rightarrow \infty$. In contrast, for the full electromagnetic case it is very difficult to fabricate materials with zero permittivity $\varepsilon$, for example; only approximate results can be achieved with lossy materials based on resonances \cite{engetha}. 

However, some advantages of the full electromagnetic case have not yet a counterpart in magnetostatics. One of these is the possibility of having negative-$\mu$ materials; whereas resonances in different kinds of natural and artifical substances can yield negative values of $\mu$ and $\varepsilon$ at non-zero frequencies \cite{veselago,pendry,padilla,freire,KSun}, no such negative-$\mu$ materials exist in magnetostatics \cite{dolgov}. Negative values of $\mu$ and $\varepsilon$ have enabled very interesting novel phenomena for electromagnetic waves, like perfect lenses \cite{perfectlens,pf2,shelby}. Some interesting physical devices such as
'illusions', in which the waves reflected by an object are made to resemble those arising from a different one \cite{illusion,illusion_static} or exterior cloaks, in which an object is cloaked at a distance \cite{exterior,wegener,guevara}, have never been experimentally realized in magnetostatics, because they require materials with negative values of $\mu$. Devising ways to create the effective response of negative-$\mu$ materials would thus pave the way towards the realization of these properties also for static magnetic fields.

In this work we 
introduce negative static permeability as a new tool for manipulating magnetic fields. By solving Maxwell magnetostatic equations for negative-permeability values, we find that a negative-permeability material can be effectively realized by a suitably tailored set of currents arranged in space. The theory is confirmed by experimentally constructing a spherical shell that emulates a negative-permeability material in a uniform magnetic field. In this way the effective properties of three-dimensional negative-$\mu$ materials can be produced in practice, overcoming the fact that passive materials with negative permeability do no exist in magnetostatics \cite{dolgov}.

The paper is structured as follows. We start in section II by studying the general magnetic response of a solid ellipsoid of isotropic permeability $\mu$, either positive or negative, to a uniform applied field. In section III we analyze hollow isotropic ellipsoids, focusing on cylindrical and spherical shells. In section IV we extend the study to the case of cylindrical and spherical shells made of anisotropic materials. In section V we discuss how magnetization currents can be used to emulate negative-permeability materials. The experimental demonstration of a negative-$\mu$ spherical shell is presented in section VI. Finally, in sections VII and VIII the obtained results and their implications are discussed, and some conclusions are extracted.

\section{Solid isotropic ellipsoids with negative permeability}

We start our study of magnetic materials with negative permeability with the case of solid bodies. For the sake of generality, we consider an ellipsoid, with semiaxes $a$, $b$ and $c$. 
The ellipsoid geometry encompasses two geometries of practical interest, the sphere ($a=b=c$) and the long cylinder in perpendicular applied field ($b=c$ and  $a\rightarrow \infty$), as well as the two other conceptually interesting cases of a long thin strip (finite $b,a\rightarrow \infty$ and $c\rightarrow 0$), and an infinite slab (finite $b$ and $a=c \rightarrow \infty$). The four geometries are shown in the insets of Fig. \ref{4dep}.

\subsection{Dependence of the fields on the permeability}

Consider an ellipsoid made of a homogeneous and isotropic material with relative magnetic permeability $\mu$, in a uniform field $\bold{H_0}$  applied along the ellipsoid $c$ axis.
The magnetic moment $\mathbf{m}$ of an ellipsoid in magnetostatics is analogous to the polarizability of a dielectric ellipsoid in a uniform applied electrostatic field. By taking into account that the magnetization is $\bold{M} = {\rm d}\bold{m}/{\rm d}V$ and the solutions in \cite{craig1983absorption}, \cite{theory} and \cite{landau}, one obtains 
\begin{equation}\label{Msolid}
\bold{M} = \frac{\mu-1}{1+N (\mu-1)} \bold{H_0}, 
\end{equation}
where $N$ is the demagnetizing factor of the ellipsoid, which  ranges from $0$ (long slab) to $1$ (thin film). For a sphere $N=1/3$ and for a long cylinder in transversal field $N=1/2$.

By using that the total magnetic field is equal to the applied one plus the demagnetizing field, $\bold{H} = \bold{H_0} + \bold{H_d}$, and that the demagnetizing field is related to the magnetization through $\bold{H_d}= - N \bold{M}$, we find that $\bold{H}$ in the ellipsoid volume can be written as $
\mathbf{H} =\mathbf{H_0}/[1+N(\mu-1)] $.
The magnetic induction is $\bold{B} = \mu_0 \left(\bold{H} + \bold{M} \right) = \mu \mu_0 \mathbf{H}$, where $\mu_0$ is the vacuum permeability.
These equations for $\mathbf{M}$, $\mathbf{H}$ and $\mathbf{B}$  show that all the fields for linear, isotropic and homogeneous ellipsoids are uniform in the material and only depend on two parameters: $\mu$ and $N$ \cite{landau,Jackson}. These equations have been considered until now only for positive values of $\mu$. However, there is in principle no physical argument against generalizing them to include negative values of $\mu$.

\begin{figure}[hbtp]
\centering
\includegraphics[scale=0.35]{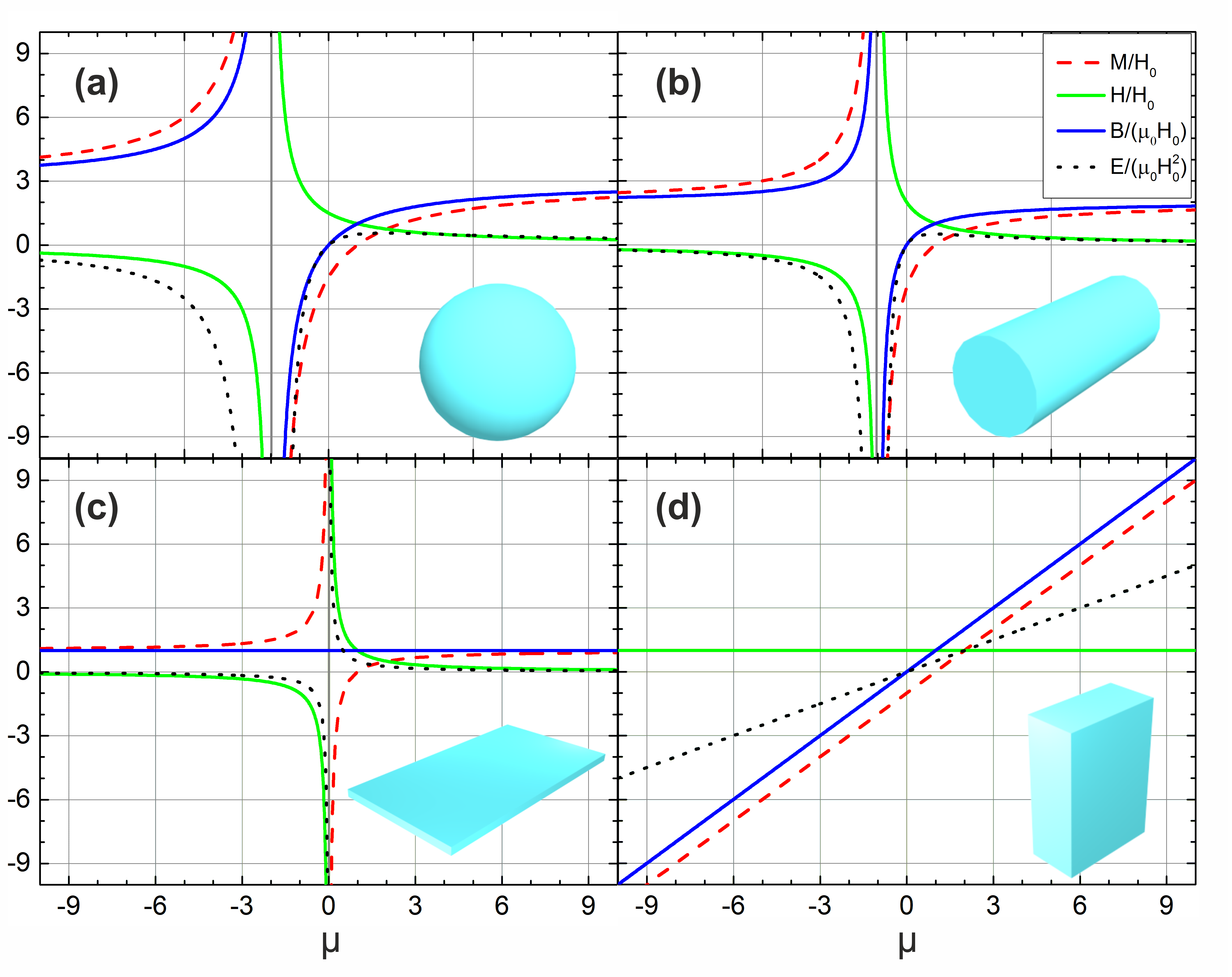}
\caption{Normalized magnetization (in red), magnetic field (in green), magnetic induction (in blue) and energy density (in black) as a function of the permeability $\mu$ for (a) a sphere, (b) an infinite cylinder in perpendicular field, (c) an infinite thin film and (d) an infinite slab.}
\label{4dep}
\end{figure}

We now describe how $\bold{M}$, $\bold{H}$, and $\bold{B}$ in the ellipsoid volume depend on $\mu$. In Fig. \ref{4dep} these fields together with the energy density $E={\bf B}\cdot{\bf H}/2$ are represented for the sphere, cylinder, thin film, and slab geometries. 
We start from the positive $\mu$ region, for which $\bold{B}$  and $\bold{H}$ have both the same direction as $\bold{H_0}$. The material is paramagnetic when $\bold{M}$ is parallel to $\bold{H_0}$  ($\mu>1$) and diamagnetic when they are antiparallel ($\mu<1$).  When $\mu=1$, $\mathbf{M} = 0$; this is the only situation in which a solid ellipsoid does not distort an applied magnetic field.
The case of $\mu\to 0$ corresponds to a perfect diamagnet (e.g. an ideal superconductor), for which $\bold{M}$ exactly cancels $\bold{H}$, resulting in ${{\bf B}(\mu= 0)} =0$. 
The hitherto unexplored regime of $\mu<0$ can be understood as a natural extension of the $\mu>0$ behavior. Decreasing $\mu$ from $\mu= 0$ one sees that, while $M$ is still negative (diamagnetic response) and ${H}$ still positive, the sign of ${B}$ becomes negative (except for the thin film case, for which strong demagnetizing effects yield $B$ independent of $\mu$).  
The absolute value of ${B}$ inside the ellipsoid continuously builds up as $\mu$ decreases from 0 to $\mu \rightarrow (N-1)/N$, where all fields diverge.
These asymptotes appear at $\mu=-2$, -1, 0, and $-\infty$ for the sphere, cylinder, thin film, and slab, respectively.
The process of increasing $\bold{B}$ (in the opposite direction to $\bold{H_0}$) with decreasing $\mu$ towards negative values is somehow symmetric to the increase of $\bold{B}$ (in the same direction as $\bold{H_0}$) observed when $\mu\rightarrow \infty$. However, the latter has a bound, $B(\mu\rightarrow \infty)=\mu_0 H_0/N$, whereas the building up of negative $\bold{B}$ eventually diverges.

When crossing the divergence, with further decrease of $\mu$, $\bold{B}$, $\bold{H}$ and $\bold{M}$ change their sign, and the ellipsoid becomes paramagnetic. 
Decreasing $\mu$ to more negative values results in a decrease of the absolute value of $\bold{B}$, $\bold{H}$ and $\bold{M}$. Interestingly, the limit $\mu \rightarrow -\infty$ corresponds to the ideal soft ferromagnetic limit, $\mu  \rightarrow \infty$.

Based on these results, the conventional concept of diamagnetic and paramagnetic responses (negative and positive $\bold{M}$, respectively) acquires a new more general meaning. For negative $\mu$, the diamagnetic and paramagnetic responses are not longer bounded, as happens for magnetic materials with positive $\mu$ [$H_0/(N-1) < M < H_0/N$ for $\mu>0$]. Instead, the diamagnetic or paramagnetic responses can now take arbitrarily large values, until eventually diverging at some particular (negative) $\mu$ value. Interestingly, the giant diamagnetic and paramagnetic responses can also be interpreted as the response of a superconducting or a soft ferromagnetic ellipsoid, respectively, with larger volume than that of the actual body.

\subsection{Energy analysis}

\begin{figure}[hbtp]
\centering
\includegraphics[scale=1]{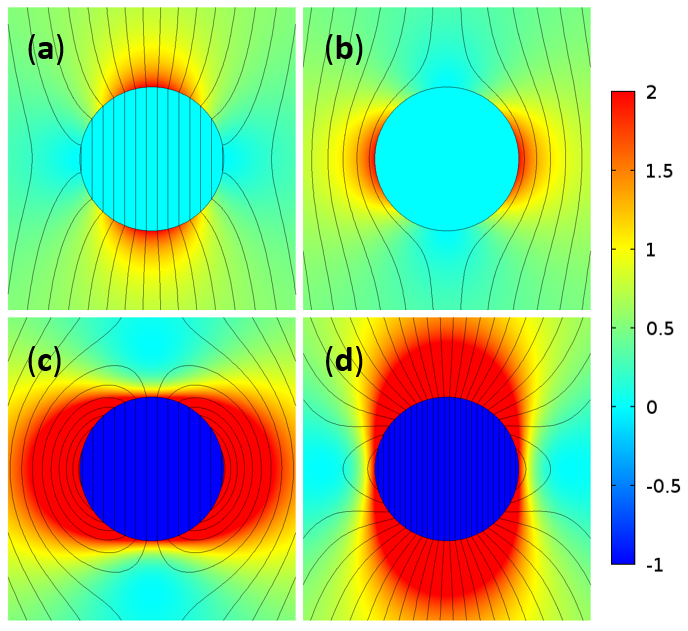}
\caption{Magnetic induction field lines and normalized energy density $E/(\mu_0 H_0^2)$ (in colors) for the magnetic response of cylinders of (a) $\mu = 10^4$, (b) $\mu = 10^{-4}$, (c) $\mu=-1/2$ and (d) $\mu=-2$ to a vertically applied magnetic field $H_0$. }
\label{prof_cylinder}
\end{figure}

In general, magnetostatic phenomena can be regarded as a spatial reorganization of magnetic energy. The two typical magnetic materials with more extreme values of $\mu$, soft ferromagnets ($\mu \rightarrow \infty$) and perfect diamagnets such as superconductors ($\mu\to 0$) expel the magnetic energy from their interior (superconductors because {\bf B}=0 and ferromagnets because {\bf H}=0). Therefore, when a uniform magnetic field is applied to a soft ferromagnetic or superconducting material, the energy is excluded from the materials volume and redistributed into the rest of space [Fig. \ref {prof_cylinder}(a) and (b)]. The same occurs for materials with intermediate $\mu>0$, in this case with only partial expulsion of energy. Materials with negative $\mu$ [Fig. \ref {prof_cylinder}(c) and (d)] expel even more energy than the $\mu \rightarrow \infty$  and $\mu\to 0$ cases. Energy balance is preserved in negative-$\mu$ materials because the extra expelled energy is compensated by negative energy in the materials. The energy density $E={\bf B}\cdot{\bf H}/2$  is negative in all the volume of a negative-$\mu$ material since $B$ and $H$ have opposite signs.

To further understand negative-$\mu$ materials, it is useful to analyze the behavior of magnetic materials in terms of  magnetic field lines. When $\mu>1$, lines are attracted towards the material, being $\mu\rightarrow\infty$   the case of maximum attraction [Fig. \ref{prof_cylinder}(a)]. When $0<\mu<1$, the effect is the opposite, field lines tend to avoid the material volume, until the perfect diamagnetic case $\mu\to 0$ is reached and all the field lines skip the cylinder [Fig. \ref{prof_cylinder}(b)]. 
With further reducing $\mu$ to negative values a double effect starts to build up: the lines from the applied field are expelled farther, as if the material was more diamagnetic than a superconductor, and at the same time some closed field lines appear [Fig. \ref{prof_cylinder}(c)].  The closed field lines generated by the material become larger as $\mu$ approaches the asymptote at $\mu\to (N-1)/N$ [$\mu=-1$ for a cylinder, as seen in Fig. \ref{4dep}].  When continuing towards more negative values of $\mu$ [Fig. \ref{prof_cylinder}(d)], the closed field loops diminish until they disappear in the limit $\mu \rightarrow -\infty$.
In magnetostatics, when dealing with linear materials as in our case, closed field lines can only arise from currents, because of Ampere's law. In Section \ref{emulating} we discuss how to find the required current distributions to emulate a negative-$\mu$ material. Because these currents will need to be readjusted when changing the applied field, negative-$\mu$ materials devised in this way can be classified as active \cite{alu_review,green,active_thermal,
active_tiejuncui}.

\subsection{Conjugate relations}
A general property of the magnetization of the ellipsoids emerges when negative values of the permeability are considered. For any ellipsoid of permeability $\mu$ there exists a conjugate ellipsoid of permeability $\mu'$, which has exactly the same magnetization with opposite sign, $
\bold{M(\mu')} = - \bold{M(\mu)}$.
By using Eq. (\ref{Msolid}), we find that the conjugate permeability of $\mu$ is
\begin{equation}\label{solidCONJ}
\mu' = 1 - \frac{\mu-1}{1+2N(\mu-1)},
\end{equation}
which only depends on the geometry of the ellipsoid.

Considering a solid cylinder, $N=1/2$,  Eq. (\ref{solidCONJ}) yields $\mu' = 1/\mu$. Then $\mu'$ has  the same sign as $\mu$. This conjugate relation was obtained for positive $\mu$ values for long rectangular bars in transversal field in \cite{Pardo2000} and also appeared in \cite{demaghollowcylinder} for long hollow cylinders in transversal field.

In the case of a solid sphere, conjugate relations have not been explored before. Using $N=1/3$, Eq. (\ref{solidCONJ}) leads to a conjugate permeability $\mu' = (-\mu+4)/(2\mu+1)$, which shows that the sign of $\mu'$ is not always the same as that of $\mu$. When $\mu$ is larger than 4, the conjugate sphere does not have a positive value of $\mu'$. For this reason conjugate relations for  spheres could not be obtained without taking into account that $\mu$ can take negative values. The conjugate $\mu'$ of a soft ferromagnetic sphere ($\mu \rightarrow \infty$), for example, is a sphere with $ \mu'= -1/2$, instead of a superconducting sphere ($\mu'$=0) as for a cylinder.

\section{Hollow isotropic ellipsoids with negative permeability}

We continue our study of negative-$\mu$ materials by considering the case of hollow bodies. Novel features such as magnetic field concentration in the hole of the bodies \cite{concentrator} appear in this geometry.

Consider an ellipsoidal homogeneous and isotropic material with relative magnetic permeability $\mu$ and semi axes $a_2$, $b_2$ and $c_2$ with a centered hole of semi axes $a_1$, $b_1$ and $c_1$; we restrict our study to the case of both the hole and the outer surface having the same shape. A uniform magnetic field, $\bold{H_0}$, is applied along a principal axis of the ellipsoid.

Different from solid bodies, the magnetic response of a hollow ellipsoid to a uniform magnetic field has in general not only a dipolar term but higher orders as well. 
Only the cases of hollow spheres and cylinders, because of their high symmetry, have a dipolar response. Analytic expressions for the magnetic fields for hollow spherical and cylindrical shells can be found in the Appendix.

The dipolar magnetic moment $\bold{m}$ of a general hollow ellipsoid is analogous to the polarizability resulting from the application of an electric field to a hollow dielectric ellipsoid \cite{craig1983absorption}. From this,
one can calculate the averaged magnetization on the whole ellipsoid volume, $V$, including the hole volume, as $\bold{M^*} =(\int_V \bold{M}(r,\theta) dV)/V$. Its expression is  
\begin{equation}\label{M}
\bold{M^*} = \frac{(f-1) \left(N(\mu-1)-\mu \right) (\mu-1)}{(f-1)(N-1)N(\mu-1)^2+\mu} \bold{H_0}, 
\end{equation}
which is uniform and in the direction of the applied magnetic field, as in the case of a solid ellipsoid. $f$ is the fraction of the external ellipsoid occupied by the hole, $f = (a_1 b_1 c_1)/(a_2 b_2 c_2)$. In the limits $\mu\to \pm \infty$, $M^*$ tends to $1/N$, as for solid bodies. When $f\to 0$ we recover Eq. (\ref{Msolid}).

Eq. (\ref{M}) shows that there are two values of $\mu$ that result in a divergence of the magnetization,
\begin{equation} \label{mu_inf1}
\mu_{1,2} = 1 + \frac{ -1 \pm \sqrt{1+4N(f+N-f N-1)}}{2(f-1)(N-1)N},
\end{equation}
Bearing in mind that 0 $\leq$ f $ < $ 1 and 0 $\leq$ $N$ $\leq$ 1 it is seen that these two values of $\mu$ are negative for any ellipsoid.

When $M^*$=0 the shell does not create a dipolar response; in the case of a hollow sphere or cylinder this makes the object magnetically undetectable because the applied magnetic field is not distorted.
Whereas for solid ellipsoids the magnetization is zero only in the trivial case of no material, $\mu_{\rm ND}$ = 1, for isotropic hollow ellipsoids there is an extra solution for $\bold{M^*} (\mu_{\rm ND})$=0. By using Eq. (\ref{M}) we find that
\begin{equation}
\mu_{\rm ND} = \frac{N}{N-1},
\end{equation}
which does not depend on $f$, but only on the geometry of the ellipsoid through its demagnetizing factor. In the particular cases of spherical and cylindrical shells, the non-distortion permeabilities are $\mu_{\rm ND} = -1/2$  and $\mu_{\rm ND} = -1$, respectively.

\begin{figure}[hbtp]
\centering 
\includegraphics[scale=0.5]{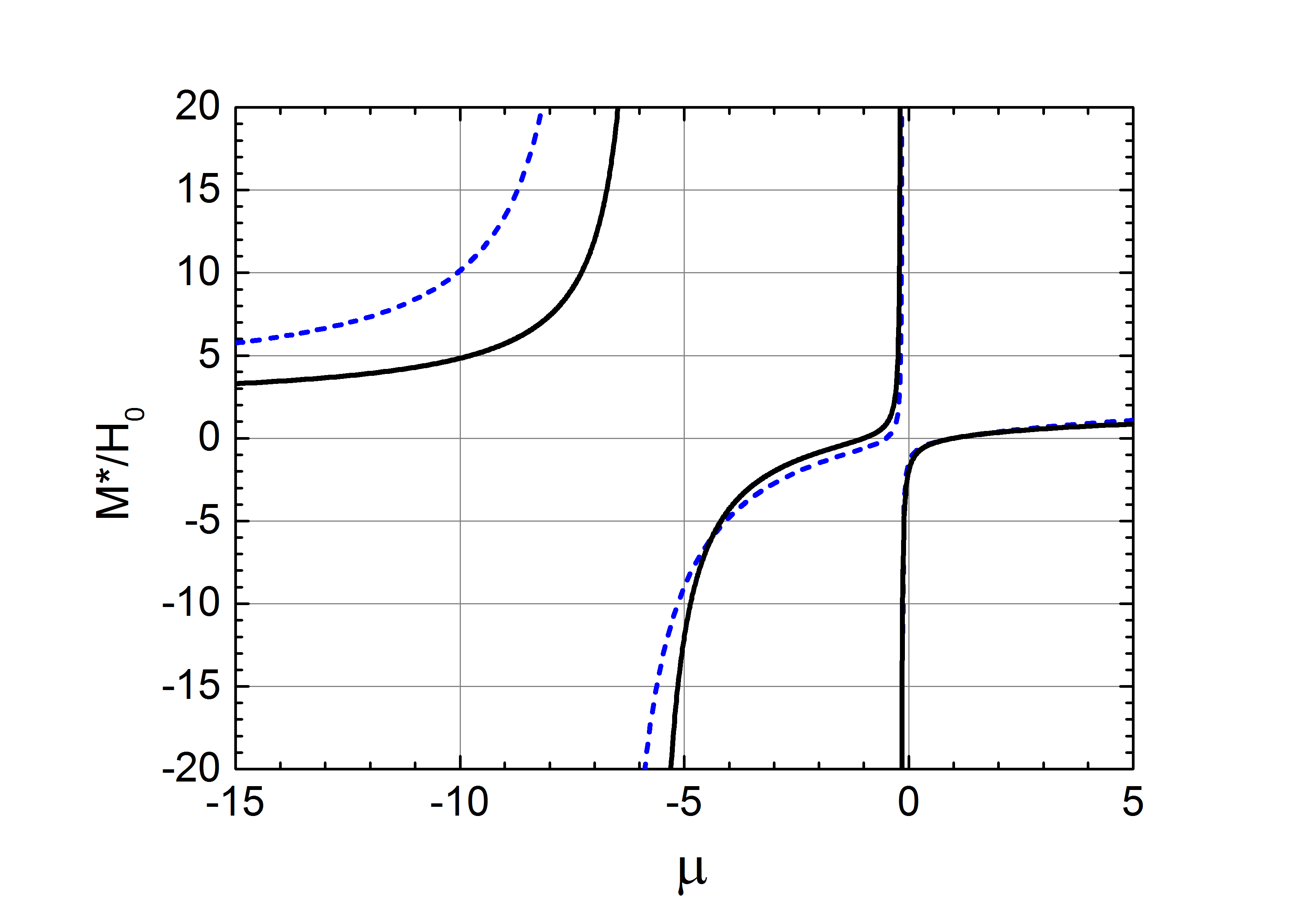}
\caption{Normalized averaged magnetization $M^*$ as a function of the permeability $\mu$ for a spherical shell (blue dashed line) and a cylindrical shell (black solid line), for $f=1/2$. \label{Mstar}}
\end{figure}

We show in Fig. \ref{Mstar} the dependence of $M^*$ upon $\mu$ for a hollow sphere ($N=1/3$) and a hollow long cylinder ($N=1/2$) for $f=1/2$, where the two divergences and the two non-distortion permeabilites can be seen, for each geometry.

\subsection{Conjugate relations}

The consideration of negative values of the permeability leads to conjugate relations for hollow ellipsoids,
as for solid ones. For any hollow ellipsoid of permeability $\mu$ two conjugate ellipsoids of permeabilities $\mu'_1$, and $\mu'_2$, exist which have exactly the same magnetization with opposite sign, $\bold{M^*}(\mu'_{1,2}) = - \bold{M^*}(\mu)$.

For a general hollow ellipsoid, the conjugate relations are found using Eq. (\ref{M}). In the particular case of a hollow cylinder the result $\mu'_1 = 1/\mu$, obtained in \cite{demaghollowcylinder} is recovered. This is the same conjugate relation that appeared for a solid cylinder. Interestingly, a new solution appears as
\begin{equation}
\mu'_2 =  \frac{(\mu-1)f + (\mu+1)}{(\mu-1)f - (\mu+1) }.
\end{equation}

Conjugate relations for a hollow sphere can also be analytically obtained through cumbersome expressions (not shown). None of them corresponds to the case of a solid sphere.

\section{Hollow Anisotropic Cylindrical and Spherical Shells}

We now continue our study of hollow bodies by considering shells with homogeneous anisotropic permeabilities, extending the results studied above for isotropic materials. We restrict the results to the two more relevant geometries, spherical and cylinidrical shells.

Consider homogeneous and anisotropic spherical and cylindrical shells of external radius $R_2$ and internal radius $R_1$, characterized by their angular and radial relative permeabilities, $\mu_\theta=\mu_\varphi$ and $\mu_r$, and $\mu_\theta$ and $\mu_\rho$, respectively.
A uniform magnetic field $\bold{H_0}$ is applied in the $z$ direction. 
Magnetostatic Maxwell equations can be analytically solved (see Appendix for the full derivation), providing the solutions for the magnetic field in the three different regions: inside the hole, in the shell and in the external region. For positive $\mu$, these solutions were studied in \cite{spherical_conc,concentrator,expconcentrator}.

The solutions show two important properties.
First, the magnetic field inside the hole of the shells is always uniform and has the direction of the applied one, ${\bf H}_s^{\rm INT} = -a_s{\bf H_0}$ and ${\bf H}_c^{\rm INT} = -a_c{\bf H_0}$, for the spherical and the cylindrical shell, respectively. The expressions for the coefficients $a_s$ and $a_c$ are shown in Eqs. \ref{a_s_an} and \ref{a_c_an}. 
Second, the magnetic field in the external region is, in general, modified with respect to the applied field due to the presence of the shell. The field created by the shell corresponds to the field created by a centered dipole with magnetic moment pointing in the applied field direction, ${\bf m}_s = 4 \pi b_s {\bf H}_0$ for a spherical shell and ${\bf m}_c = 2 \pi b_c {\bf H}_0$ for a cylindrical one. The expressions for the coefficients $b_s$ and $b_c$ are shown in Eqs. \ref{b_s_an} and \ref{b_c_an}. 
A positive (negative) value of $b_s$ or $b_c$ indicates that the shell is paramagnetic (diamagnetic).

\begin{figure}[hbtp]
\centering
\includegraphics[scale=0.17]{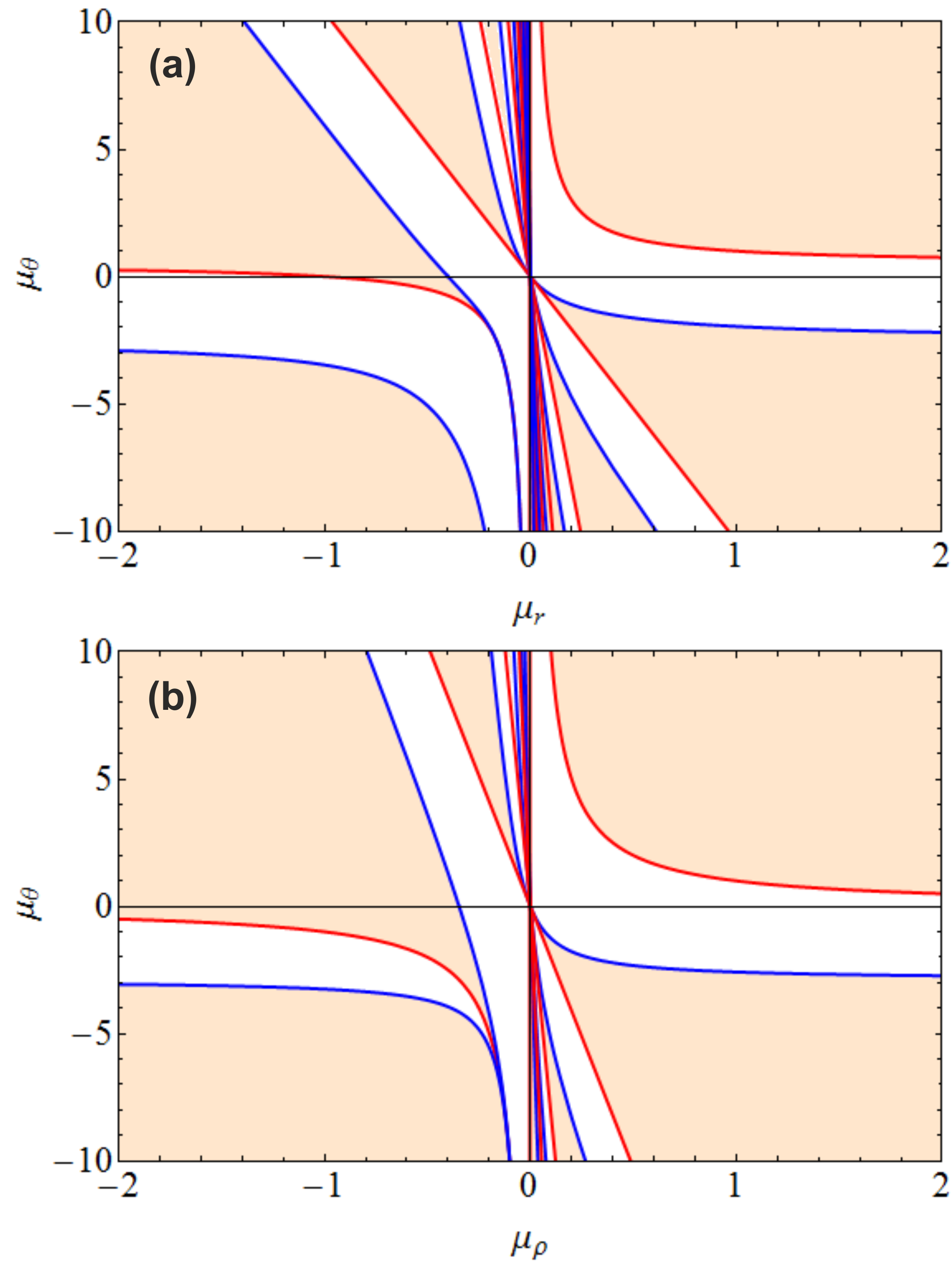}
\caption{Relations 
of non-distortion (red lines) and divergent fields (blue lines)
between the permeabilities (a) $\mu_\theta$ and $\mu_r$ for a spherical shell and (b) $\mu_\theta$ and $\mu_\rho$ for a cylindrical one. $R_2/R_1 =2$ for both cases. The regions filled in orange correspond to paramagnetic shells, while the white regions correspond to diamagnetic shells. \label{figmuTvsmuResfera3}}
\end{figure}

In the following we analyze the anisotropic shells that do not distort a uniform applied magnetic field as well as those that involve divergent magnetic fields. 
The overall results can be seen in Fig. \ref{figmuTvsmuResfera3}, where the permeability relations resulting in a non-distorting shell and those leading to divergent fields are plotted. It is seen that these two cases alternate, so that there is always a line of no distortion between two consecutive lines of field divergence. Also, these lines constitute the borders between diamagnetic and paramagnetic regions. In this way, the concept of paramagnetic and diamagnetic materials is enriched. For conventional materials with positive $\mu$ (right upper quadrant in Fig. \ref{figmuTvsmuResfera3}) there is a single frontier line separating the two regions. In the general picture that negative $\mu$ is bringing, the border lines and the paramagnetic and diamagnetic regions increase until reaching an infinite number of them.

\subsection{Non-distortion shells}
It can be obtained from Eqs. (\ref{b_s_an}) and (\ref{b_c_an}) that for a given radial permeability there are infinite values of the angular permeability for which the coefficients $b_s$ and $b_c$ become zero, and thus the shells do not distort the external magnetic field. They can be grouped into two types of solutions. For a spherical shell,
\begin{align}
\mu_\theta &= \frac{1+\mu_r}{2 \mu_r}, \label{nodistesfera_an}\\
\mu_\theta &= -\frac{\mu_r}{8} \left[ \left( \frac{2 \pi n}{{\rm ln}(R_2/R_1)}\right)^2 +1 \right], \quad n=1,2,3..., \label{nodist+-esfera}
\end{align}

and for a cylindrical shell,
\begin{align}
\mu_\theta &= \frac{1}{\mu_\rho}, \label{nodistcilindre_an}\\
\mu_\theta &= - \mu_\rho \left( \frac{ \pi n}{{\rm ln}(R_2/R_1)}\right)^2
 , \quad n=1,2,3... . \label{nodist+-}
\end{align}

The first type of solutions [Eqs. (\ref{nodistesfera_an}) and (\ref{nodistcilindre_an})] corresponds to the red curves in  
Fig. \ref{figmuTvsmuResfera3}, extending mainly in the first and third quadrant. These solutions were explored for positive $\mu$ in \cite{spherical_conc,concentrator}
The non-distorting isotropic shells studied above, $\mu_r = \mu_\theta =-1/2$ and $\mu_\rho = \mu_\theta = -1$ for a spherical and a cylindrical shell, respectively, are particular cases of these solutions. The second type [Eqs. (\ref{nodist+-esfera}) and (\ref{nodist+-})] corresponds to the red straight lines in  
Fig. \ref{figmuTvsmuResfera3}, extending in the second and fourth quadrants. There is an infinite number of these lines, and their slope depends upon a single parameter, $n$.

\subsection{Magnetic field concentration inside the hole of a non-distorting shell}

The two types of non-distorting solutions of Eqs. (\ref{nodistesfera_an})-(\ref{nodist+-}) differ in the field concentrated inside their hole.

We start studying the first type of solutions [Eqs. (\ref{nodistesfera_an}) and (\ref{nodistcilindre_an})]. The field in the hole for a spherical and a cylindrical shell is, respectively,
\begin{align}
H_s^{\rm INT}=H_0(R_2/R_1)^{1-1/\mu_r}, \label{Eq:H_s_CONC}\\
H_c^{\rm INT}=H_0(R_2/R_1)^{1-1/\mu_\rho},
\label{Eq:H_c_CONC}
\end{align}
where we have used Eqs. (\ref{a_s_an}) and (\ref{a_c_an}).

The magnetic field concentration can be interpreted in terms of energy reorganization. Since we are considering shells that do not distort the external field, the energy density in the external region is the same as if there was no shell. When the permeabilites are positive, the concentration of energy inside the hole [$E^{\rm INT} = \mu_0 (H^{\rm INT})^2/2$] can be simply understood considering that part of the energy that was in the space occupied by the shell has been redistributed and placed inside the hole.  
When permeabilities are negative, the minimum concentration occurs for an infinitely large negative radial permeability and is $H^{\rm INT}_{\rm  min} = \left(R_2/R_1 \right)H_0$, independently of the shell geometry. Interestingly, this corresponds to the maximum concentration that can be achieved with positive permeabilities, occurring when the radial permeability tends to $+\infty$. When the radial permeability approaches $0^-$ the field concentration increases, and diverges in this limit.

\begin{figure}[hbtp]
\centering
\includegraphics[scale=0.32]{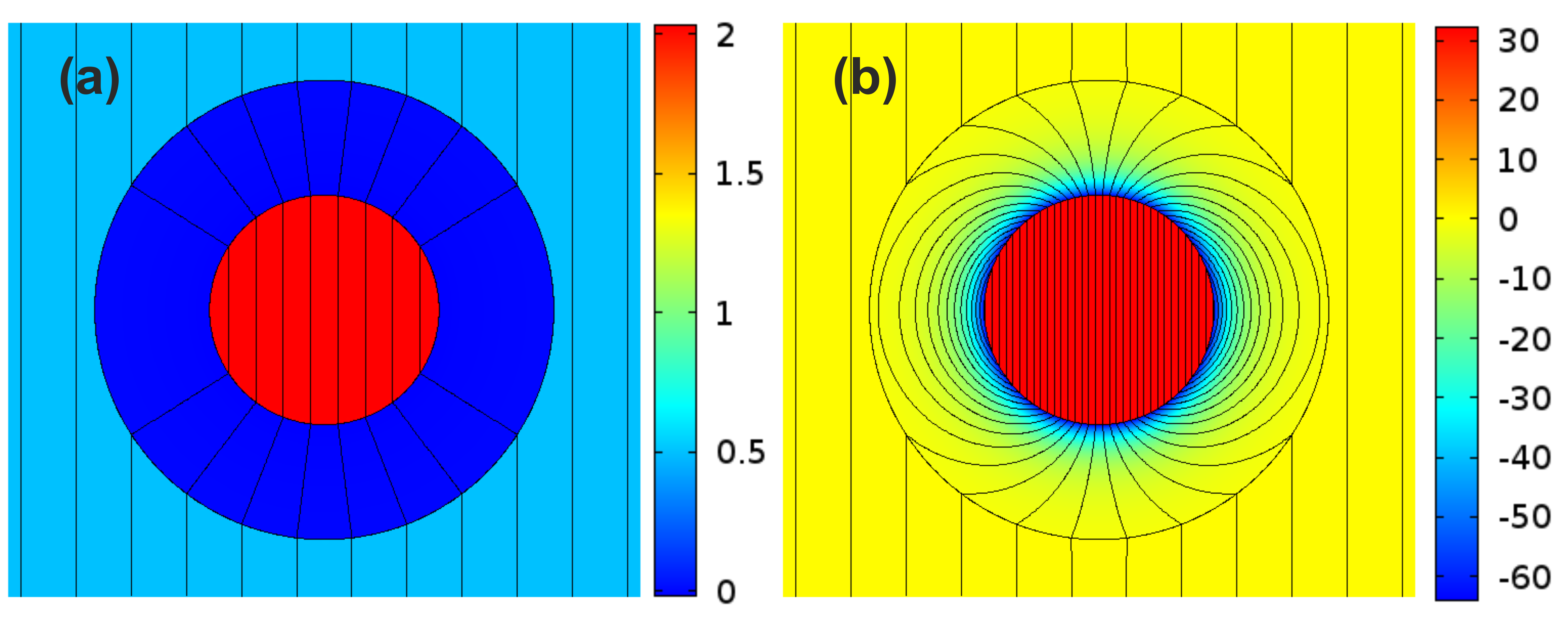}
\caption{Magnetic induction field lines and normalized energy density $E/(\mu_0H^2_0)$ in color scale for two cylindrical shells with radii ratio $R_2/R_1$ = 2 and magnetic permeabilities (a) $\mu_\rho =100$ and $\mu_\theta =0.01$ and (b) $\mu_\rho = -1/2$ and $\mu_\theta =-2$. \label{F_nodistcilindre_an}}
\end{figure}

To explain how this large magnetic field concentration is achieved we compare the behaviour of two non-distorting shells fulfilling the non-distortion relation of Eq. \eqref{nodistcilindre_an}, one with positive $\mu$ and the other one with negative $\mu$, for the cylindrical geometry (Fig. \ref{F_nodistcilindre_an}). When $\mu >0$, the energy density inside the hole, $E^{\rm INT}$, is maximum when the energy density in the shell is zero. This happens when $\mu_\rho \rightarrow \infty$ and $\mu_\theta \rightarrow 0$ [in Fig. \ref{F_nodistcilindre_an}(a) this is approximated by $\mu_\rho =100$ and $\mu_\theta = 0.01$]. In this situation, all the energy that was in the space occupied by the shell has been redistributed and placed inside the shell hole. When considering a shell with negative $\mu$ [Fig. \ref{F_nodistcilindre_an}(b)] the energy inside the hole $E^{\rm INT}$ is larger than that for positive $\mu$. Since the energy in the external region is the same for both cases, energy conservation requires that the energy in a negative-$\mu$ shell volume is negative, as shown in Fig. \ref{F_nodistcilindre_an}(b).

\begin{figure}[hbtp]
\centering
\includegraphics[scale=0.32]{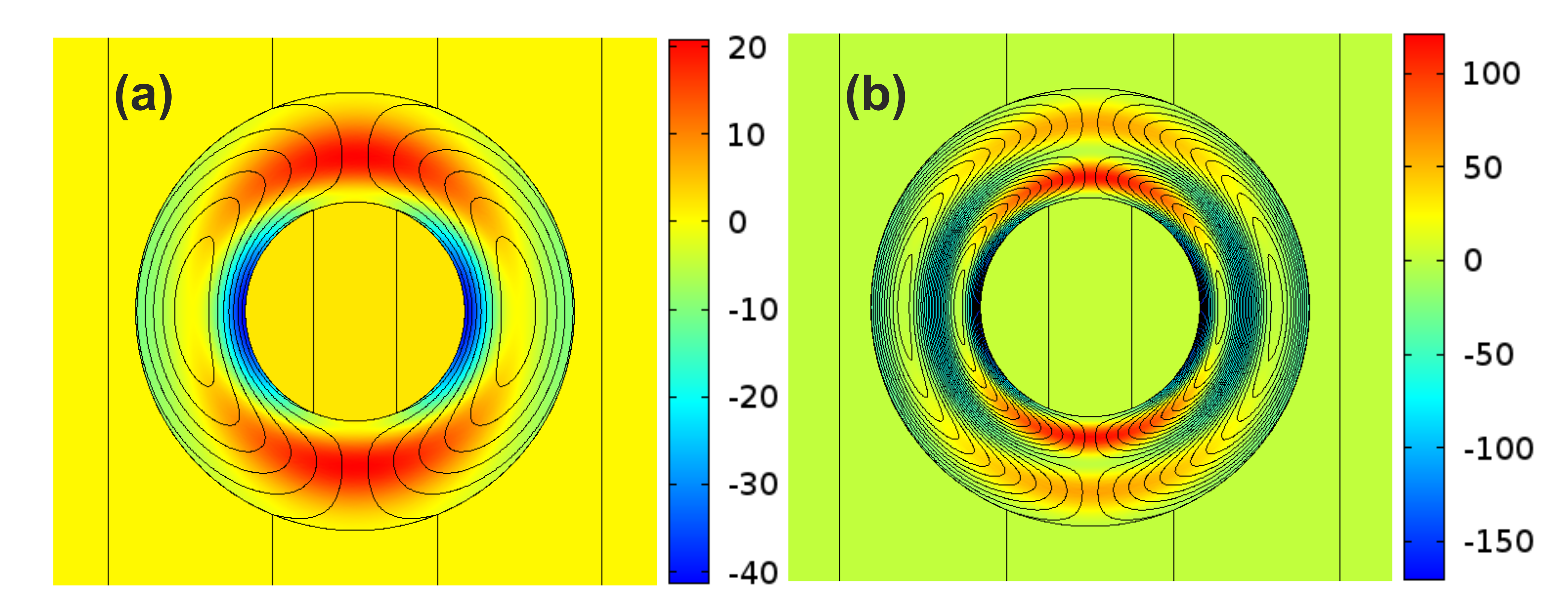}
\caption{Magnetic induction field lines and normalized energy density $E/(\mu_0H_0^2)$ in color scale for two cylindrical shells with radii ratio $R_2/R_1$ = 2. Both have $\mu_\rho = 1$ and their corresponding $\mu_\theta$ is obtained from Eq. \eqref{nodist+-} for (a) $n$=1 and (b) $n$=2. \label{figuranodist+-}}
\end{figure}

Now we analyze the field concentration corresponding to the second type of non-distortion solutions, resulting from Eqs. \eqref{nodist+-esfera} and \eqref{nodist+-}. Interestingly, for all shells fulfilling these equations the field inside the hole is $H_{\rm s}^{\rm INT} = \pm H_0 (R_2/R_1)^{3/2}$ for a spherical shell and $H_{\rm c}^{\rm INT} = \pm H_0 (R_2/R_1)$ for a cylindrical shell, according to Eq. \eqref{a_s_an} and \eqref{a_c_an}, respectively; the sign is positive when $n$ is even and negative when $n$ is odd. Therefore, the energy density inside the hole is the same for all the solutions of this type. This is illustrated in the examples of Fig. \ref{figuranodist+-}, where the energy density and the magnetic field lines are represented for two cylindrical shells. It is seen that $n$ indicates the number of regions inside the shell that are surrounded by closed magnetic field lines. 

\subsection{Divergences of fields}
The permeability relations yielding divergent fields  can be found from the zeroes in the denominators of Eqs. (\ref{a_s_an}) and (\ref{a_c_an}). It is interesting that when $\alpha^2$ and $k^2$ are negative, there are an infinite number of such relations. The divergences occur, for spherical and cylindrical shells, respectively, when 
\begin{equation}
\sqrt{-\alpha^2} {\rm ln}(R_2/R_1) = 2{\rm arctan} \left(\frac{3\mu_r \sqrt{-\alpha^2}}{\beta}\right) + 2\pi n, \quad  \label{inf_esfera}
\end{equation}

\begin{align}
\sqrt{-k^2} {\rm ln}(R_2/R_1) &= {\rm arctan}\left({\frac{-2 \mu_\rho \sqrt{-k^2}}{\mu_\rho \mu_\theta+1}}\right) + \pi n, \label{inf_cilindre}
\end{align}
where $n$=0,1,2.... These expressions are represented as blue lines in Fig. \ref{figmuTvsmuResfera3} for a particular shell with $R_2/R_1=2$.

\section{Emulating negative-permeability materials}
\label{emulating}

Media with negative magnetic permeability do not exist in magnetostatics, as demonstrated in \cite{dolgov}. However, we next show how negative-$\mu$ media can be effectively emulated by replacing them with a set of currents. In order to find these currents we use the general property that in magnetostatics the magnetic response of a material can be obtained by substituting it with its magnetization currents. Given an arbitrary magnetic material in an applied magnetic field, $\mathbf{H}_{\rm 0}$, the corresponding surface and volume magnetization currents can be calculated from the magnetization of the material, $\mathbf{M}$, respectively, as
\begin{align}
\mathbf{K}_{\rm M}&=\mathbf{M}\times \mathbf{n},  \label{K}
\\
\mathbf{J}_{\rm M}&=\nabla \times \mathbf{M}, \label{J} 
\end{align}
where $\mathbf{n}$ is a unitary vector perpendicular to the material surface.

The total magnetic induction in all the space (even at points inside the material), $\mathbf{B}$, can be simply calculated as the applied magnetic induction, $\mathbf{B}_0=\mu_0 \, \mathbf{H}_0$, plus the magnetic induction created by these magnetization currents, $\mathbf{B}_{\rm c}$. 
Therefore, by externally supplying the adequate set of currents the total distribution of $\mathbf{B}$ will be exactly the same as if that material was present. This allows to emulate any magnetic material, even materials with negative permeabilities.

We next find the currents emulating a negative-$\mu$ material in the case of a spherical shell, which is the one we will experimentally demonstrate below. 
Consider a spherical shell with inner and outer radii $R_1$ and $R_2$, respectively, and homogeneous relative magnetic permeabilities $\mu_r$, $\mu_{\theta}$, and $\mu_{\varphi}$. Its response to a uniform magnetic field {\bf H}$_0$ applied in the $z$ direction is analytically obtained (see Appendix). Restricting to isotropic materials $\mu_r=\mu_{\theta} \equiv \mu$ ($\mu_{\varphi}$ is irrelevant due to the symmetry of the applied field), the corresponding magnetization currents are calculated from Eqs. (\ref{K}) and (\ref{J}) taking into account that, by definition, {\bf M}=$(\mu-1)${\bf H} and {\bf H} in the material region can be obtained from Eq. (\ref{HSHEesfera_an}), as
\begin{equation}
\label{Eq:Km1}
\mathbf{K}_{\rm M} (r=R_1)= \dfrac{-18\mu(\mu -1) \left(R_2/R_1\right)^3 H_0{\rm sin}\theta} {-4(\mu-1)^2+(4\mu ^2+10\mu +4)\left(R_2/R_1\right)^3}
\mathbf{e_\varphi},
\end{equation}

\begin{equation}
\label{Eq:Km2}
\mathbf{K}_{\rm M}  (r=R_2) = \dfrac{6(\mu -1)\left[(\mu -1)+(2\mu +1)\left(R_2/R_1\right)^3\right] H_0{\rm sin}\theta}{-4(\mu-1)^2+(4\mu ^2+10\mu +4)\left(R_2/R_1\right)^3}\mathbf{e_\varphi},
\end{equation}
\begin{equation} \label{Eq:Jm}
\mathbf{J}_{\rm M} = 0.
\end{equation}
Since we consider homogeneous and isotropic materials, no volume magnetization currents appear.

\section{Experimental demonstration of a negative-permeability material}

We now experimentally demonstrate our theoretical ideas and the plausibility of emulating magnetic materials with negative $\mu$. We consider a homogeneous and isotropic spherical shell with inner and outer radii $R_1$ and $R_2$, respectively. 
We choose a permeability $\mu=-0.5$; this shell does not distort the applied field and concentrates the field in the hole by a factor $(R_2/R_1)^3$ [Eq. (\ref{Eq:H_s_CONC})]. These properties cannot be simultaneously obtained by conventional materials with positive $\mu$.

\subsection{Emulation of a negative-permeability material by a finite set of currents}

\begin{figure}[htb]
	\centering
		\includegraphics[width=0.6\textwidth]{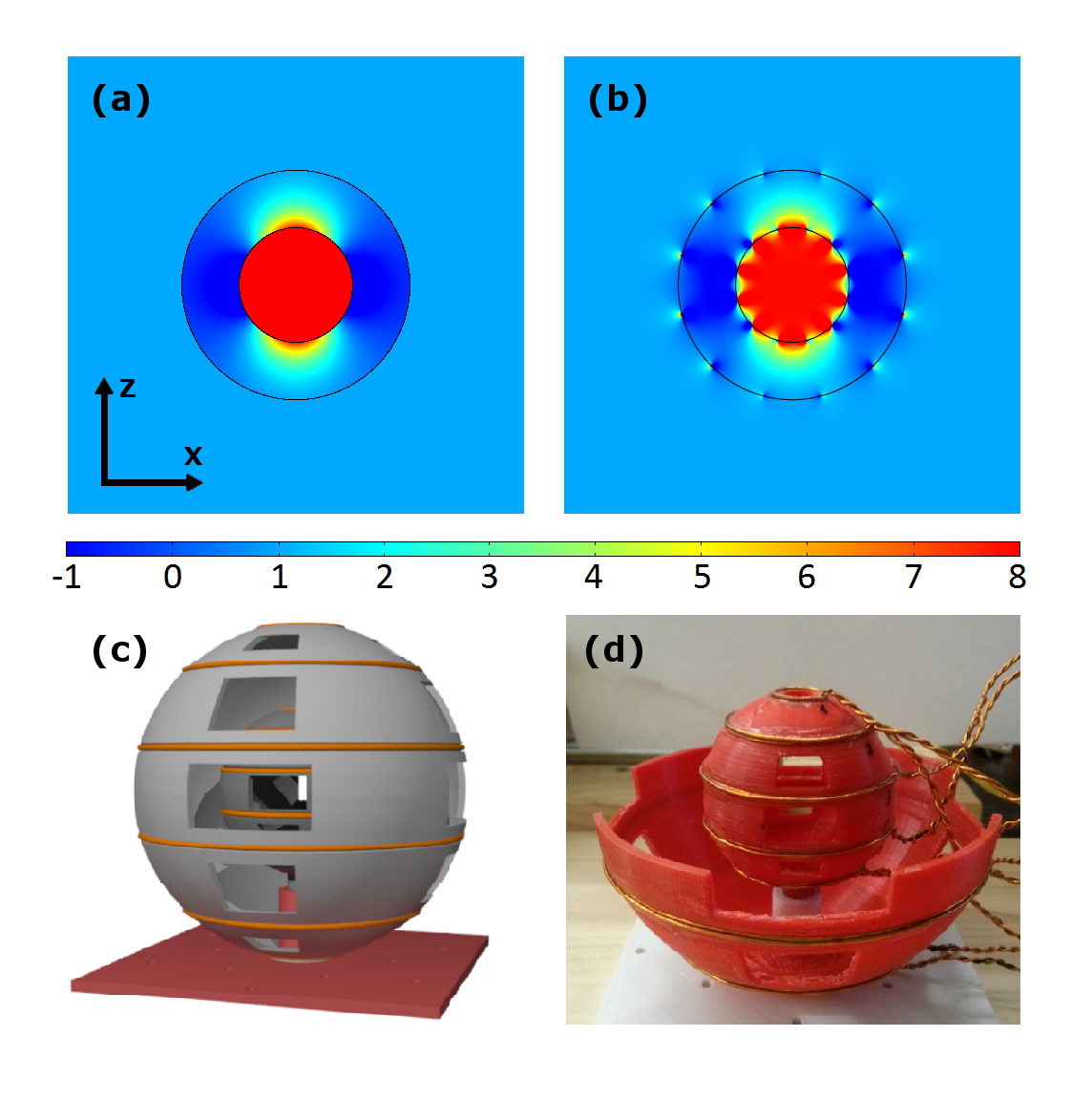}
	\caption{(a) Finite-element simulation of the $z$-component of {\bf B}, normalized to $B_0$, when a field $B_0$ is applied in the $z$-direction to a spherical shell with $\mu=-0.5$ and radii ratio $R_2/R_1=2$. (b) Same for the discretized 6+6 current loops. (c)
3D sketch of the experimental negative-$\mu$ material, consisting of two sets of 6 circular current loops placed on a 3D-printed plastic former. (d) Picture of the actual experimental negative-$\mu$ material.
} 
	\label{construction}
\end{figure}

To construct an actual spherical shell with effective negative permeability $\mu=-0.5$, the surface currents given by Eqs. \eqref{Eq:Km1} and \eqref{Eq:Km2} have to be externally supplied at the inner and outer surfaces of the shell, respectively. These continuous current distributions are converted into discrete sets of current loops in our practical realization. 
Numerical simulations (by the AC/DC module of the Comsol Multiphysics software) indicate that the discretization into 6 current loops at each of the surfaces [Fig. \ref{construction}(b)] approximates reasonably well the field created by the theoretical continuous current distribution [Fig. \ref{construction}(a)].
The current corresponding to each loop is calculated as the integral of the surface current, 
\begin{align}
I(R_a,\theta_i)=\int^{\theta_i+\pi/ 12}_{\theta_i-\pi/ 12} K_{{\rm M}}(r=R_a) \, R_a \, {\rm d}\theta, \label{discre}
\end{align}
where $\theta_i$ is the angular position of each current loop and $a=1,2$.

\subsection{Experimental setup and feedback loop}
\label{feedback}

\begin{figure}[htb]
	\centering
		\includegraphics[width=0.6\textwidth]{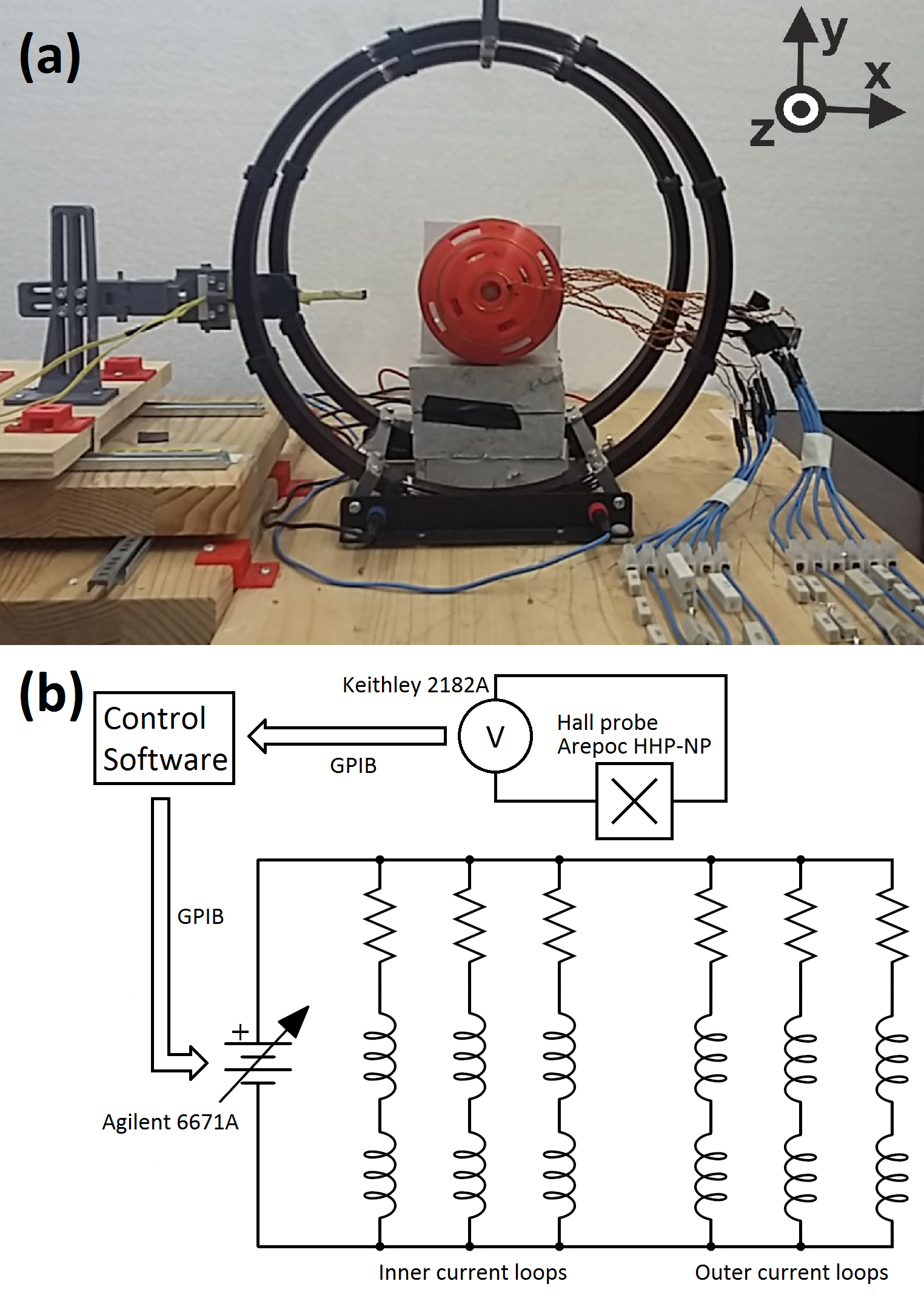}
	\caption{(a)Picture of the experimental setup with the spherical negative-$\mu$ material in the middle of two Helmholtz coils that create a uniform field in the $z$ direction; the tip of the Hall probe is shown on the left of the sphere. (b) Scheme of the feedback loop circuit.} 
	\label{experiments_feedback}
\end{figure}

In our experiments, the 6+6 current loops, each consisting of 3 turns of copper wire, are placed onto two specially designed 3D-printed spherical formers, with radii $R_1$=25mm and $R_2$=50mm, respectively [Figs. \ref{construction}(c) and \ref{construction}(d)].
The spherical shell is placed in between a pair of Helmholtz coils, which create a uniform magnetic field in the $z$ direction in the sphere region, as shown in Fig. \ref{experiments_feedback}(a).

For a given applied field value, the required currents at each loop can be obtained from Eqs. (\ref{Eq:Km1}), (\ref{Eq:Km2}) and (\ref{discre}). They are fed in the 12 loops using a common voltage source from a Agilent 6671A power supply; each loop is connected in series with a load resistor, whose value is calculated to provide the required current.

If the applied field is changed, the value of the current in the loops needs to be readjusted in order to keep emulating the same negative-$\mu$ material. For this purpose we setup a feedback loop that automatically adjusts the currents to the applied field value [Fig. \ref{experiments_feedback}(b)].

For the feedback loop we use a LabView Virtual Instrument as a Control Software, with a process described as follows. First, the applied field is measured with a Hall probe. Then, the currents corresponding to the reading of the field value are calculated according to Eqs. (\ref{Eq:Km1}), (\ref{Eq:Km2}) and (\ref{discre}). Finally, these currents are fed into the loops by using the same resistors and the Control Software automatically readjusts the input voltage. Thanks to the linear dependence between the current and the field and the simplicity of the experimental components, the feedback loop is very robust against possible instabilities arising from fluctuations of the measured applied field.

In this way, we achieve an effective negative-$\mu$ material. Even though the feedback loop mechanism is theoretically valid for any applied field value, in practice the range of applicability is limited by the power dissipation of the resistors and the overall available input power. The discretization we have used is adequate for uniform magnetic fields applied perpendicular to the loops, but the general procedure could be adapted to different field distributions using other discretization schemes.

\begin{figure}[htb]
	\centering
		\includegraphics[width=0.6\textwidth]{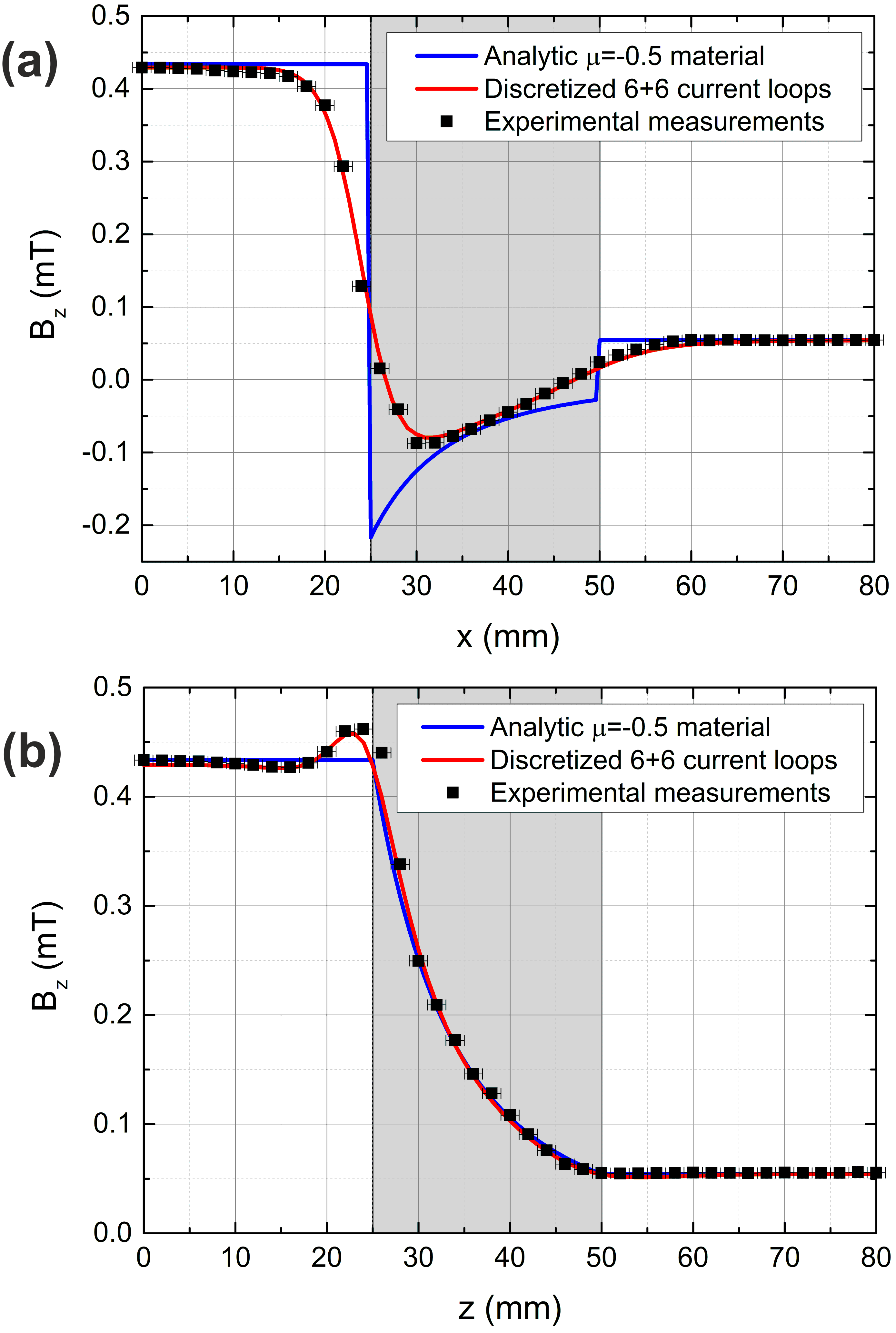}
	\caption{(a) Experimental measurements (black squares), finite-element calculation for the discretized spherical shell with $\mu=-0.5$ (red line), and analytic results for the ideal material (blue line) for the $z$-component of {\bf B} along the $x$-axis. (b) Same as (a) for the $z$-component of {\bf B} along the $z$-axis. The shell region is painted in grey color.} 
	\label{experiments}
\end{figure}

\subsection{Field measurements}

In order to verify that the actual device acts as a material with $\mu=-0.5$, we apply a  
magnetic induction $B_0=\mu_0H_0$=0.0543mT and compare the measured field profiles with the theoretical results. The $z$ component of the magnetic induction is measured with a Hall probe along the $x$ [Fig.  \ref{experiments}(a)] and $z$ [Fig. \ref{experiments}(b)] directions. The experimental results show that the field inside the hole is uniform and that the external field is not modified by the presence of the shell, verifying the theory. The field in all regions coincides very well with the numerical simulations of the discretized device. Only close to the surfaces there is a small discrepancy between the ideal and the discretized cases because of the discretization.

\section{Discussion}

Negative properties of materials are an intense recent topic of research in physics, including negative acoustic \cite{xie,guen}, negative mechanical properties \cite{negmec,negmec2,negmec3} and negative capacitance \cite{negcap,negcap2}. Most of these systems are very complicate to realize in practice. In contrast, the negative-$\mu$ materials we introduce in this work can be simply realized by a set of suitably tailored electrical currents whose analytic expressions are found. These currents are  proportional to the uniform applied magnetic field. Therefore, to emulate the response of a particular negative-$\mu$ material, one first has to sense the applied magnetic field and then set the required currents. Because of this sensing-setting requirement we can regard the proposed negative-$\mu$ magnetic materials as active. The feedback loop presented in Section \ref{feedback} automatically adapts the currents to the magnetic field, allowing the emulation of a negative-$\mu$ material even when the applied field is changed.

Having negative-$\mu$ magnetostatic materials may enable a whole new set of possibilities for controlling magnetic fields, analogous to those proposed or demonstrated for the full electromagnetic case. One of the most dramatic
properties enabled by materials with negative refraction index is achieving illusions, that is, objects that appear as different objects when illuminated by light \cite{illusion}. In \cite{companion} we demonstrate how to obtain illusion in magnetostatics using negative-$\mu$ materials. The magnetic signature of a magnetic material (a soft ferromagnet in \cite{companion}) is transformed into that of a different one (a perfect diamagnet) by enclosing the former in a shell emulating a negative-$\mu$ behavior. Other illusions such as magnifying or shrinking materials, cloaks, and anticloaks \cite{castaldi,chenluo,switchable,
superscatterer,transformation} can also be realized using the same scheme \cite{companion}.
Another intriguing possibility that may eventually become possible based on our results may be the realization of exterior cloaks \cite{exterior,wegener,guevara}. As stated by Wegener in \cite{wegener} conventional 
metamaterial cloak needs to be wrapped around the object, so it would be yet more stunning and useful if it could rather be spatially separated from the object. Such exterior cloaking has been demonstrated experimentally in dc electrical conduction using effectively negative electric conductivities in a plane \cite{FangYan} of active metamaterials \cite{Brownian}. Our results open the door to construct a magnetic cloak that can act at a distance in a full 3D scheme, something which may have applications in many areas involving magnetic fields, such as medical imaging techniques.

\section{Conclusions}

We have
introduced materials with negative static permeability as a new tool for manipulating magnetic fields. 
We have explored solutions of Maxwell magnetostatic equations considering negative-$\mu$ materials. A whole new set of solutions have emerged, extending those previously known for the conventional case of positive $\mu$ materials. For solid ellipsoid bodies, which include the physically interesting cases of a sphere and a cylinder in perpendicular field, the consideration of negative $\mu$ brings the existence of a divergence of magnetic fields at a particular negative-$\mu$ value, which only depends on the body demagnetizing factor. For hollow isotropic cylinders and spheres with negative $\mu$, there are two values of $\mu$ for which fields diverge, and also an extra solution for cloaking magnetic fields, apart from the trivial solution of no material, $\mu=1$. Some conjugate relations between the magnetic responses of bodies of different permeabilities have been found, bringing to light some hidden symmetries that become apparent when considering the case of negative $\mu$.
For cylindrical and spherical shells with anisotropic permeability new families of solutions arise, including an infinite number of cloaking situations, of divergent  magnetic fields, and also of infinite borders between paramagnetic and diamagnetic regions (at which the magnetization of the body changes from positive to negative, respectively). 
For all studied cases, magnetization currents can be obtained from the analytic expressions of the field distributions.
We have demonstrated that negative-permeability materials can be realized in practice by replacing the material with these magnetization currents. We have experimentally confirmed these ideas by constructing a set of current loops that emulates the properties of a spherical shell with $\mu = -0.5$. Our theoretical results and the emulation of negative-$\mu$ materials by currents may create new ways of controlling magnetic fields.

\section*{ACKNOWLEDGEMENTS}

We thank European Union Horizon 2020 Project FET-OPEN MaQSens (grant agreement 736943), and
projects MAT2016-79426-P (Agencia Estatal de Investigación / Fondo Europeo de Desarrollo Regional) and 2014-SGR-150
for financial support.  A. S. acknowledges a grant from ICREA Academia, funded by the Generalitat de Catalunya. 

\renewcommand{\theequation}{A\arabic{equation}}
\setcounter{equation}{0}

\section*{APPENDIX: ANALYTIC EXPRESSIONS FOR SPHERICAL AND CYLINDRICAL SHELLS}
Consider homogeneous, linear, and anisotropic spherical and cylindrical shells of external radius $R_2$ and internal radius $R_1$, with an applied magnetic field $\bold{H_0}$ in the z direction. The angular and radial relative permeabilities are $\mu_\theta=\mu_\varphi$ and $\mu_r$ for the spherical shell and $\mu_\theta$ and $\mu_\rho$ for the cylindrical shell. Since there are no free currents in the system, \textbf{$\nabla$} $\times$ \textbf{H}=0, and the magnetic field can be written in terms of a magnetic scalar potential $\phi$, \textbf{H} = -\textbf{$\nabla$}$\phi$, in all the space. Using this equation and knowing that \textbf{$\nabla$} $\cdot$ \textbf{B} = 0, the magnetic field in the three different regions: inside the hole (INT), in the shell (SHE) and in the external region (EXT) can be obtained. For a spherical shell,

\begin{align}
\mathbf{H_{\rm s}^{\rm INT}} (r, \theta) &= H_0\left[ -a_{\rm s} {\rm cos}\theta \mathbf{e_r} + a_{\rm s} {\rm sin}\theta \mathbf{e_\theta}\right],\label{HINTesfera}\\
\mathbf{H_{\rm s}^{\rm SHE}} (r, \theta)  &= H_0\left[ \left(\frac{(1-\alpha )c_{\rm s}}{2 r^{(3-\alpha)/2}} +
\frac{d_{\rm s}(1+\alpha )}{2 r^{(3+\alpha)/2}}\right) {\rm cos}\theta \mathbf{e_r}  + \left( \frac{c_{\rm s}}{r^{(3 -\alpha)/2}}+ \frac{d_{\rm s}}{r^{(3+\alpha)/2}}\right) {\rm sin}\theta\mathbf{e_\theta}\right] ,   \label{HSHEesfera_an}\\
\mathbf{H_{\rm s}^{\rm EXT}}  (r, \theta) &= H_0\left[ \left(\frac{2 b_{\rm s}}{r^3} +1\right) {\rm cos}\theta\mathbf{e_r} + \left(\frac{b_{\rm s}}{r^3} -1\right) {\rm sin}\theta \mathbf{e_\theta}\right], \label{HEXTesfera}
\end{align}
and for a cylindrical shell,

\begin{align}
\mathbf{H_{\rm c}^{\rm INT}} (\rho, \theta) &= H_0\left[-a_{\rm c}  {\rm cos}\theta \mathbf{e_\rho} + a_{\rm c} {\rm sin}\theta\mathbf{e_\theta}\right],\label{HINTcilindre}\\
\mathbf{H_{\rm c}^{\rm SHE}} (\rho, \theta) &=H_0\left[  \left(-c_{\rm c}k\rho^{k-1} + \frac{d_{\rm c} k }{\rho^{k+1}}\right) {\rm cos}\theta \mathbf{e_\rho} + \left(c_{\rm c} \rho^{k-1} + \frac{d_{\rm c}}{\rho^{k+1}}\right) {\rm sin}\theta \mathbf{e_\theta}\right] ,\label{HSHEcilindre_an}\\
\mathbf{H_{\rm c}^{\rm EXT}} (\rho, \theta) &= H_0\left[ \left(\frac{b_{\rm c}}{\rho^2}+1\right) {\rm cos}\theta \mathbf{e_\rho}  + \left(\frac{b_{\rm c}}{\rho^2}-1\right) {\rm sin}\theta\mathbf{e_\theta}\right]. \label{HEXTcilindre}
\end{align}
where e have used $\alpha^2 = 8\mu_\theta/\mu_r +1$ and $k^2=\mu_\theta/\mu_\rho$.

The coefficients of the magnetic field can be obtained by applying the boundary conditions (continuity of radial component of {\bf B} and tangencial component of {\bf H} at both surfaces $R_1$ and $R_2$). For a spherical shell,

\begin{align} 
a_{\rm s} &= \frac{6 \mu_r \alpha \left(R_2/R_1\right)^{(3+ \alpha)/2}}{\beta - 3\mu_r\alpha - (\beta + 3\mu_r\alpha)\left(R_2/R_1\right)^\alpha}, \label{a_s_an}\\
c_{\rm s} &=  \frac{3 (\mu_r\alpha + \mu_r +2)R_2^{(3+\alpha)/2}  R_1 ^{-\alpha}}{\beta - 3\mu_r\alpha - (\beta + 3\mu_r\alpha)\left(R_2/R_1\right)^\alpha},\label{c_s_an}\\
d_{\rm s} &= \frac{3 (\mu_r\alpha - \mu_r  -2) R_2^{(3+\alpha)/2} }{\beta - 3\mu_r\alpha - (\beta + 3\mu_r\alpha)\left(R_2/R_1\right)^\alpha}, \label{d_s_an}\\
b_{\rm s} &= \frac{-2( 2\mu_r\mu_\theta- \mu_r -1)\left[\left(R_2/R_1\right)^{\alpha} - 1\right]R_2^3 }{\beta - 3\mu_r\alpha - (\beta + 3\mu_r\alpha)\left(R_2/R_1\right)^\alpha},\label{b_s_an}
\end{align}
where $
\beta=4\mu_r\mu_\theta+\mu_r +4$. For a cylindrical shell,
\begin{align} 
a_{\rm c} &= \frac{4\mu_\rho k \left(R_2/R_1\right)^{1+k} }{(\mu_\rho k-1 )^2-(\mu_\rho k+1 )^2\left(R_2/R_1\right)^{2k}}, \label{a_c_an}\\
c_{\rm c} &= \frac{2 (\mu_\rho k+1 )R_2^{1-k} \left(R_2/R_1\right)^{2k} }{(\mu_\rho k-1 )^2-(\mu_\rho k+1 )^2\left(R_2/R_1\right)^{2k}},\label{c_c_an}\\
d_{\rm c} &= \frac{2(\mu_\rho k -1){R_2}^{1+k} }{(\mu_\rho k-1 )^2-(\mu_\rho k+1 )^2\left(R_2/R_1\right)^{2k}}, \label{d_c_an}\\
b_{\rm c} &= \frac{-(\mu_\rho\mu_\theta-1)  R_2^2 \left[ \left(R_2/R_1\right)^{2k}-1\right] }{(\mu_\rho k-1 )^2-(\mu_\rho k+1 )^2\left(R_2/R_1\right)^{2k}}.\label{b_c_an}
\end{align}

From Eqs. (\ref{HEXTesfera}) and (\ref{HEXTcilindre}) the magnetic field in the exterior region is, in general, modified with respect to the applied field due to the presence of the shell. The field created by the shell corresponds to the field created by a dipole with magnetic moment $m_s = 4 \pi b_s$ or $m_c = 2 \pi b_c$, for a spherical and a cylindrical shell, respectively. Eqs. (\ref{HINTesfera}) and (\ref{HINTcilindre}) show that the magnetic field inside the hole of these shells is always a uniform field aligned in the same or in the opposite direction to $\bold{H_0}$, with magnitude $H_z = -a$. 
Finally, Eqs. (\ref{HSHEesfera_an}) and (\ref{HSHEcilindre_an}) indicate that the magnetic field inside the material is the sum of a uniform field of magnitude $-c$ aligned in the same or in the opposite direction to $\bold{H_0}$ and a field that corresponds to the field created by a dipole of magnetic moment $4 \pi d_s$ for a spherical shell and $2 \pi d_c$ for a cylindrical shell.

\end{document}